\documentclass[11pt]{article}
\usepackage{graphicx}
\usepackage{xcolor}
\usepackage{blindtext}
\usepackage{multicol, caption}
\usepackage[margin={1.6cm, 2cm}]{geometry}
\usepackage{cite}
\usepackage{float}
\usepackage{epstopdf}
\newcommand\pubnumber{TOP-22-009}
\newcommand\pubdate{January 15, 2024}

\def\institute{Department of Physics\\
Vrije Universiteit Brussel}
\def\authemail{\footnote{Contact: jusong@cern.ch}}

\def\Title#1{\begin{center} {\Large #1 } \end{center}}
\def\Author#1{\begin{center}{ \sc #1} \end{center}}
\def\Address#1{\begin{center}{ \it #1} \end{center}}

\newcommand\pubblock{\rightline{\begin{tabular}{l} \pubnumber\\
         \pubdate  \end{tabular}}}
\newenvironment{Presented}{\begin{quotation} \begin{center} 
             PRESENTED AT\end{center}%\bigskip 
      \begin{center}\begin{large}}{\end{large}\end{center} \end{quotation}}

\def\beq{\begin{equation}}
\def\eeq#1{\label{#1}\end{equation}}
\def\eeqn{\end{equation}}

\def\beqa{\begin{eqnarray}}
\def\eeqa#1{\label{#1}\end{eqnarray}}
\def\eeqan{\end{eqnarray}}

\let\bar=\overbar

\def\Dslash{\not{\hbox{\kern-4pt $D$}}}
\def\dslash{\not{\hbox{\kern-2pt $\del$}}}

\def\msb{{\bar{\ssstyle M \kern -1pt S}}}

\def\ttbb{t\bar{t}b\bar{b}}
\def\ttH{t\bar{t}H}
\def\tt{t\bar{t}}
\def\bb{b\bar{b}}
\def\pt{p_T}
\def\dR{\Delta R}

\begin{document}
%\begin{titlepage}
\pubblock

%\vfill
\Title{DNN-based identification of additional b jets \\for a differential $\ttbb$ cross section measurement}
%\vfill
\Author{Juhee Song\authemail}
\Address{\institute}
\vfill
\textbf{Abstract -- }In differential measurements of the $\ttbb$ process, observables related to the b jets not originating from top quark decays are of special interest to probe the multi-scale QCD nature of the $\ttbb$ process, and the description of the additional b jet radiation via different simulation tools. In order to access these additional b jets, identification algorithms have to be applied to determine which of the b jets recorded in the detector are the b jets of interest. This is achieved via a DNN-based method which will be highlighted in this article. 
\vfill

\begin{Presented}
$16^\mathrm{th}$ International Workshop on Top Quark Physics\\
(Top2023), 24--29 September, 2023
\end{Presented}
\vfill

\def\thefootnote{\fnsymbol{footnote}}
\setcounter{footnote}{0}

\begin{multicols}{2}

\section{Introduction}
\begin{figure}[H]
    \centering
    \includegraphics[width=0.7\linewidth]{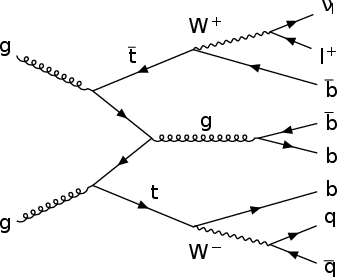}
    \caption{$\ttbb$ process in semi-leptonic channel. Each top quark decays into a W boson and a b quark. Subsequently, the W boson can decay either hadronically into a quark and an anti-quark pair or leptonically into one charged lepton and a neutrino. In the semi-leptonic decaying mode specifically, one W boson decays hadronically, resulting in a quark and anti-quark pair, while the other W boson decays leptonically, producing one charged lepton and a neutrino. In addition to the b quarks originating from the top quarks, there are two more b quarks resulting from gluon splitting, which are additional b quarks.}
    \label{fig:ttbb}
\end{figure}
Modelling the associated production of top quark and bottom quark pairs, $\ttbb$, in proton-proton (pp) collisions at the LHC, is infamously challenging because of non-negligible mass of the bottom quark, and multi-scale nature involving top and bottom quarks~\cite{Buccioni:2019plc, Jezo:2018yaf}.
Therefore a significant test of perturbative quantum chromodynamics (QCD) calculations is achieved by comparing predictions for $\ttbb$ production with inclusive and differential cross section measurements.
Moreover, $\ttbb$ process is irreducible important background for searches and other measurements, including the measurement of the associated production of top quark pairs and Higgs bosons ($\ttH$), where the Higgs boson decays to a pair of bottom quarks ($H\rightarrow\bb$)~\cite{Aad:2016zqi, Aaboud:2017rss, Aaboud:2018urx, Sirunyan:2018ygk, Sirunyan:2018hoz, Sirunyan:2018mvw}, and the search for the simultaneous of four top quarks ($\tt\tt$)~\cite{Aaboud:2018jsj, ATLAS:2020hpj, ATLAS:2021kqb, Khachatryan:2014sca, Sirunyan:2017roi, Sirunyan:2017tep, CMS:2019jsc, CMS:2019rvj, CMS:2023zdh, ATLAS:2023ajo, CMS:TOP-22-013}. 
These two processes provide direct access to the top quark Yukawa coupling, a crucial parameter of the standard model (SM)~\cite{Cao:2016wib, Cao:2019ygh}.

In order to improve understanding of $\ttbb$ production measurement of $\ttbb$ inclusive and differential cross section measurements are presented in~\cite{CMS:TOP-22-009} in the CMS experiment.
For the differential cross section measurement, there are two different approaches are presented.
In the first approach, b jets are considered produced from the gluon splitting if they have the smallest angular separation.
There is no attempt to use the ancient history to identify additional b jets.
In the second approach, the b jets not from the top quark decays are identified at the generator level with a direct access to the decaying history, and a multivariate algorithm is developed to identify the resulting reconstructed b jets among all observed jets.

In this article, we focus on the second approach, aiming to identify two b jets not originating from top quark decays. 
The implementation of a multivariate algorithm enhances the precision in pinpointing additional b jets within $\ttbb$ events. 
These selected b jets are then employed to measure the differential cross section.

\section{Event samples and selections}
Monte Carlo (MC) simulated samples for $\ttbb$ events in the years 2016, 2017, and 2018 with the CMS detector, are employed to train the multivariate algorithm in this analysis. 
Various modelling approaches are applied to $\ttbb$ samples, leading to a wide range of predictions with substantial uncertainties. 
The nominal $\ttbb$ signal is generated using the four-flavour scheme (4FS), where the matrix elements for $\ttbb$ are calculated with massive b quarks. 
Additionally, the signal process is derived from the $\tt$ simulation, assuming massless b quarks, referred to as the five-flavour scheme (5FS). 
In the 5FS, additional b jets are accounted for by the Parton Shower (PS).

We use $\tt$ events, where W bosons decay semi-leptonically.
In the final state, therefore, qualified events have exactly one charged lepton and six jets, with a minimum of four of them being b jets.
Considering these criteria, the following event selections are applied. 

\begin{itemize}
    \item Exactly one isolated muon or electron
    \item At least six jets
    \item At least four b-tagged jets
\end{itemize}

\section{Analysis Strategy}
\subsection{Event-by-event approach}
We adopt an event-by-event approach to identify additional b jets at the reconstructed level. 
The definition of an additional b jet involves an angular separation ($\dR$) criterion, where the separation between the b jet and those not originating from top quarks on the generator level should be smaller than 0.4. 
The selection process involves choosing four b-tagged jets, sorting them in descending order based on $\pt$, and considering six possible combinations to select two from the four b-tagged jets.
Each combination is then assigned to a specific category, as detailed in Table~\ref{tab:jet_perm}. 
Categories 0 to 5 represent matchable events where two b jets are correctly identified as additional b jets. 
Conversely, category number 6 denotes non-matchable events, which have only one assigned jet or none.

\begin{table}[H]
 \centering
    \begin{tabular}[0.5\linewidth]{|c|c|c|c||c|}
        \hline
        Jet1 & Jet2 & Jet3 & Jet4 & \textcolor{blue}{Category} \\ \hline\hline
        \textcolor{red}{1} & \textcolor{red}{1} & 0 & 0 & \textcolor{blue}{0} \\ \hline
        \textcolor{red}{1} & 0 & \textcolor{red}{1} & 0 & \textcolor{blue}{1} \\ \hline
        \textcolor{red}{1} & 0 & 0 & \textcolor{red}{1} & \textcolor{blue}{2} \\ \hline
        0 & \textcolor{red}{1} & \textcolor{red}{1} & 0 & \textcolor{blue}{3} \\ \hline
        0 & \textcolor{red}{1} & 0 & \textcolor{red}{1} & \textcolor{blue}{4} \\ \hline
        0 & 0 & \textcolor{red}{1} & \textcolor{red}{1} & \textcolor{blue}{5} \\ \hline
        \multicolumn{4}{|c||}{Other combinations} & \textcolor{blue}{6} \\ \hline
    \end{tabular}
\caption{Jet combination categories. Jets are sorted by $\pt$, and their order is indicated by the numbers next to 'Jet,' with 'Jet1' having the highest $\pt$ among them. A red-colored '1' indicates that the corresponding jet is considered as an additional b jet, having satisfied the angular separation criteria. For instance, in an event where the first and second jets meet the criteria for being additional b jets, it falls into category 0.}
\label{tab:jet_perm}
\end{table}

\subsection{Deep neural network algorithm}
\begin{figure}[H]
    \centering
    \includegraphics[width=0.9\linewidth]{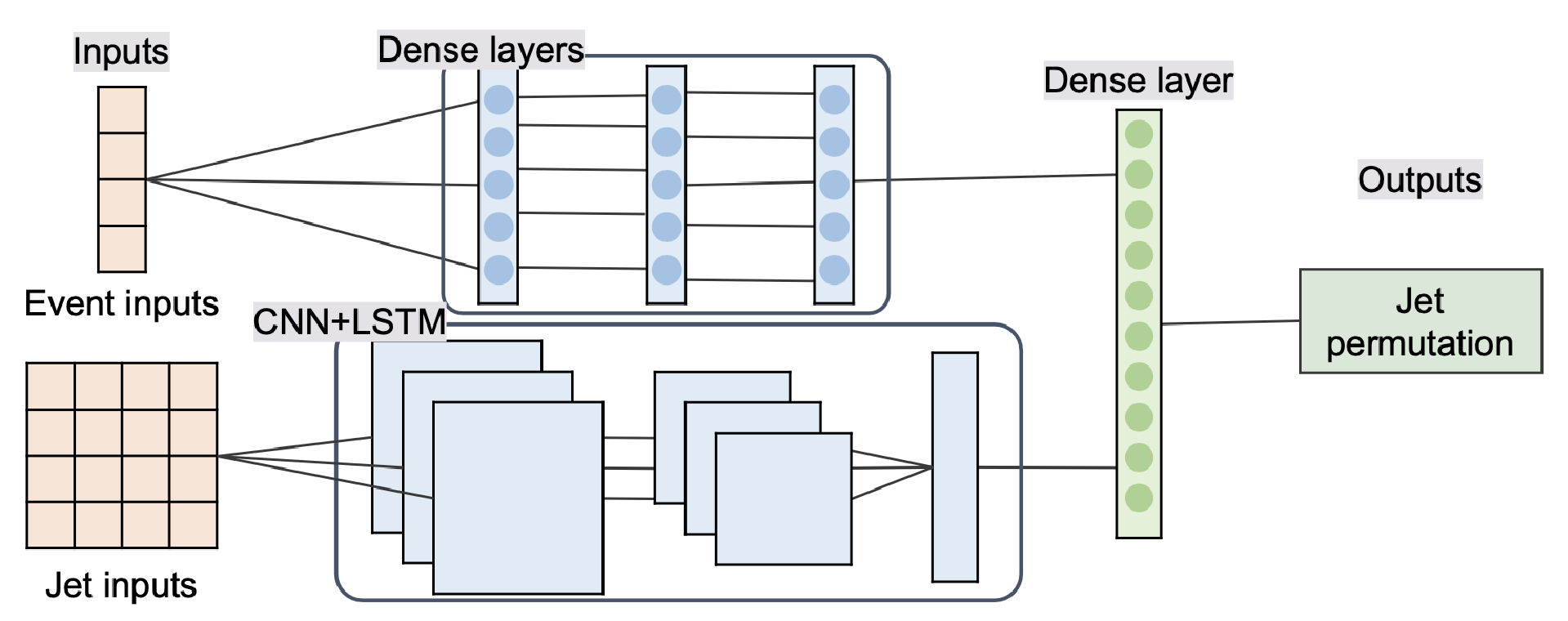}
    \caption{Structure of the neural network used for the assignment of the additional b jet pair. The network utilizes two sets of input variables: global event information is conneted to three dense network layers, and jet-specific information is connected through CNN layers and a LSTM cell. The input sequences are concatenated into one dense layer, and the output layer comprises six nodes, each representing one of the six possible candidate jet combinations.}
    \label{fig:dnn}
\end{figure}

Table~\ref{tab:jet_perm} illustrates the multiple categories that need to be assigned, making it a multi-classification problem. 
To address this, a multivariate algorithm based on a deep neural network (DNN) is employed to identify the pair of b-tagged jets most consistent with the true additional b jets indicated by the assigned categories. Figure~\ref{fig:dnn} provides an overview of the DNN architecture, inspired by the DeepJet DNN algorithm structure~\cite{Bols:2020bkb}.

The DNN utilizes two sets of input variables, specifically targeting jet-specific input information and global event information. 
Twenty jet-specific input variables are fed into convolutional neural network (CNN) layers~\cite{LeCun:1989nips}. 
These layers are followed by a long short-term memory (LSTM) cell~\cite{Hochreiter:1997yld}, enabling the network to learn correlations among jet features. 
The five features used for each candidate jet include $\pt$, $\eta$, a flag indicating whether it passes the tight b tagging working point, the angular separation ($\dR$) with the charged lepton, and the invariant mass with the charged lepton.

For global event information, thirty input variables include properties of the six dijet combinations of the four candidate b jets and the lepton. 
These input variables are connected via three dense network layers, and the input sequences are consolidated into one dense layer. 
The output layer comprises six nodes, each representing one of the six possible candidate jet combinations.
All input features are the following:
\begin{itemize}
    \item $\pt, \eta$ of each jet (2x4)
    \item flag if jet is b tagged at tight WP (1x4)
    \item $\dR$ between each jet and the lepton (1x4)
    \item invariant mass of each jet and the lepton (1x4)
    \item jet multiplicity, b-tagged jet multiplicity (2)
    \item scalar $\pt$ sum of four candidate jets (1)
    \item $\pt, \eta, \phi$ of the lepton (3)
    \item $\Delta \phi$ of dijet permutations of four candidate jets (6)
    \item $\Delta \eta$ of dijet permutations of four candidate jets (6)
    \item invariant mass of dijet permutations of four candidate jets (6)
    \item $\dR$ of dijet permutations of four candidate jets and the lepton (6)
\end{itemize}

To avoid bias in the DNN evaluation, events from the $\tt$ 5FS simulation are used for training, while the measurement evaluation utilizes the nominal 4FS $\ttbb$ sample. The DNN's performance has been validated to be independent of the choice between these simulated ttbb event samples for both training and evaluation. This ensures that measurements of DNN-based observables remain unbiased by the choice of the signal model used during DNN training.

\section{Results}
The DNN's performance is assessed in the simulation based on its ability to correctly identify the pair of additional b jets (efficiency). The DNN demonstrates an efficiency of approximately $49\%$, showcasing a notable improvement in identification accuracy compared to the 'minimum $\dR$' method, which achieves an accuracy of about 41\%. The choice of the 'minimum $\dR$' method as our benchmark is motivated by the fact that additional b jets frequently exhibit close proximity. Therefore, the 'minimum $\dR$' approach, which selects the closest two b jets, often manages to identify the correct pair.

\begin{figure}[H]
\centering
\includegraphics[width=0.95\linewidth]{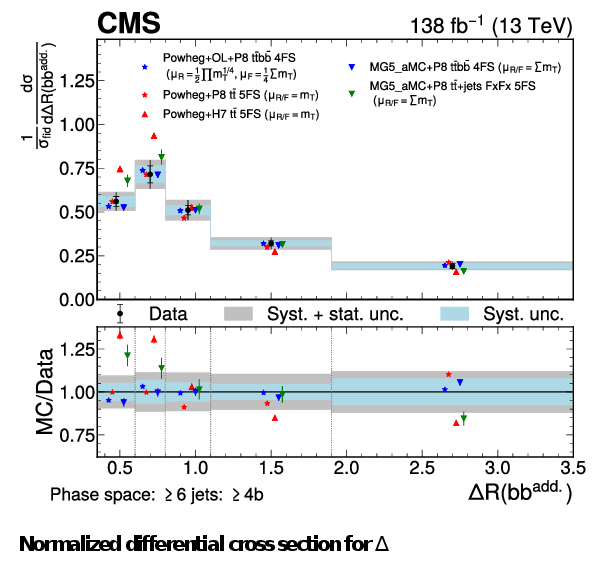}
\caption{Predicted and observed normalized $\ttbb$ differential cross sections based on $\dR$ of the additional b jet pair selected by DNN. The data are represented by points, with inner (outer) vertical bars indicating the systematic (total) uncertainties, also represented as blue (grey) bands. Cross section predictions from different simulation approaches are shown, including their statistical uncertainties, as coloured symbols.}
\label{fig:xsec}
\end{figure}

This algorithm is used to define eight observables for which differential cross section measurement is performed in ~\cite{CMS:TOP-22-009}.
Figure~\ref{fig:xsec} illustrates normalized differential cross sections of $\ttbb$ process with $\dR$ between two additional b jets selected by the DNN.
The measurements are compared to five cross section predictions of the $\ttbb$ process produced with the different combinations of event generators and PSs
The $\dR$ distribution shows a trend towards lower $\dR$ values than what is observed in data for POWHEG+H7 $\tt$ 5FS, and MG5$\_$aMC+P8 $\tt$+jets FxFx 5FS, while the other generator setups better describe the distribution.

The DNN approach offers an advantage by providing direct access to b jets not originating from top quark decays in the differential cross-section results. However, it comes with certain limitations. The analysis reveals significant migrations between the reconstruction and generator level definitions of observables, impacting the accuracy of the results. 
To address this challenge, fewer bins are used in the differential cross-section measurement.

%\Acknowledgements
%I am grateful.

\bibliography{eprint}{}

\begin{thebibliography}{10}

\bibitem{Buccioni:2019plc}
Federico Buccioni, Stefan Kallweit, Stefano Pozzorini, and Max~F. Zoller.
\newblock {NLO QCD} predictions for $\ttbb$ production in association with a light jet at the {LHC}.
\newblock {\em JHEP}, 12:015, 2019.

\bibitem{Jezo:2018yaf}
Tom{\'a}{\v{s}} Je{\v{z}}o, Jonas~M. Lindert, Niccolo Moretti, and Stefano Pozzorini.
\newblock New {NLOPS} predictions for {$\tt+b$}-jet production at the {LHC}.
\newblock {\em Eur. Phys. J. C}, 78:502, 2018.

\bibitem{Aad:2016zqi}
G.~Aad et~al.
\newblock Search for the standard model {Higgs} boson decaying into $\bb$ produced in association with top quarks decaying hadronically in {$pp$} collisions at {$\sqrt{s}=8TeV$} with the {ATLAS} detector.
\newblock {\em JHEP}, 05:160, 2016.

\bibitem{Aaboud:2017rss}
M.~Aaboud et~al.
\newblock Search for the standard model {Higgs} boson produced in association with top quarks and decaying into a $\bb$ pair in $pp$ collisions at {$\sqrt{s}=13TeV$} with the {ATLAS} detector.
\newblock {\em Phys. Rev. D}, 97:072016, 2018.

\bibitem{Aaboud:2018urx}
M.~Aaboud et~al.
\newblock Observation of {Higgs} boson production in association with a top quark pair at the {LHC} with the {ATLAS} detector.
\newblock {\em Phys. Lett. B}, 784:173, 2018.

\bibitem{Sirunyan:2018ygk}
A.~M. Sirunyan et~al.
\newblock Search for {$\ttH$} production in the all-jet final state in proton-proton collisions at {$\sqrt{s}=13TeV$}.
\newblock {\em JHEP}, 06:101, 2018.

\bibitem{Sirunyan:2018hoz}
A.~M. Sirunyan et~al.
\newblock Observation of {$\ttH$} production.
\newblock {\em Phys. Rev. Lett.}, 120:231801, 2018.

\bibitem{Sirunyan:2018mvw}
A.~M. Sirunyan et~al.
\newblock Search for {$\ttH$} production in the {$H\rightarrow\bb$} decay channel with leptonic $\tt$ decays in proton-proton collisions at {$\sqrt{s}=13TeV$}.
\newblock {\em JHEP}, 03:026, 2019.

\bibitem{Aaboud:2018jsj}
M.~Aaboud et~al.
\newblock Search for four-top-quark production in the single-lepton and opposite-sign dilepton final states in {$pp$} collisions at {$\sqrt{s}=13TeV$} with the {ATLAS} detector.
\newblock {\em Phys. Rev. D}, 99:052009, 2019.

\bibitem{ATLAS:2020hpj}
G.~Aad et~al.
\newblock Evidence for $\tt\tt$ production in the multilepton final state in proton-proton collisions at $\sqrt{s}={13TeV}$ with the {ATLAS} detector.
\newblock {\em Eur. Phys. J. C}, 80:1085, 2020.

\bibitem{ATLAS:2021kqb}
G.~Aad et~al.
\newblock Measurement of the $\tt\tt$ production cross section in {$pp$} collisions at $\sqrt{s}={13TeV}$ with the {ATLAS} detector.
\newblock {\em JHEP}, 11:118, 2021.

\bibitem{Khachatryan:2014sca}
V.~Khachatryan et~al.
\newblock Search for standard model production of four top quarks in the lepton+jets channel in {$pp$} collisions at {$\sqrt{s}=8TeV$}.
\newblock {\em JHEP}, 11:154, 2014.

\bibitem{Sirunyan:2017roi}
A.~M. Sirunyan et~al.
\newblock Search for standard model production of four top quarks with same-sign and multilepton final states in proton-proton collisions at {$\sqrt{s}=13TeV$}.
\newblock {\em Eur. Phys. J. C}, 78:140, 2018.

\bibitem{Sirunyan:2017tep}
A.~M. Sirunyan et~al.
\newblock Search for standard model production of four top quarks in proton-proton collisions at {$\sqrt{s}=13TeV$}.
\newblock {\em Phys. Lett. B}, 772:336, 2017.

\bibitem{CMS:2019jsc}
A.~M. Sirunyan et~al.
\newblock Search for the production of four top quarks in the single-lepton and opposite-sign dilepton final states in proton-proton collisions at $\sqrt{s}={13TeV}$.
\newblock {\em JHEP}, 11:082, 2019.

\bibitem{CMS:2019rvj}
A.~M. Sirunyan et~al.
\newblock Search for production of four top quarks in final states with same-sign or multiple leptons in proton-proton collisions at $\sqrt{s}={13TeV}$.
\newblock {\em Eur. Phys. J. C}, 80:75, 2020.

\bibitem{CMS:2023zdh}
A.~Tumasyan et~al.
\newblock Evidence for four-top quark production in proton-proton collisions at $\sqrt{s}={13TeV}$.
\newblock {\em Phys. Lett. B}, 844:138076, 2023.

\bibitem{ATLAS:2023ajo}
G.~Aad et~al.
\newblock Observation of four-top-quark production in the multilepton final state with the {ATLAS} detector.
\newblock {\em Eur. Phys. J. C}, 83:496, 2023.

\bibitem{CMS:TOP-22-013}
A.~Hayrapetyan et~al.
\newblock {Observation of four top quark production in proton-proton collisions at s=13TeV}.
\newblock {\em Phys. Lett. B}, 847:138290, 2023.

\bibitem{Cao:2016wib}
Qing-Hong Cao, Shao-Long Chen, and Yandong Liu.
\newblock Probing {Higgs} width and top quark {Yukawa} coupling from {$\ttH$} and $\tt\tt$ productions.
\newblock {\em Phys. Rev. D}, 95:053004, 2017.

\bibitem{Cao:2019ygh}
Qing-Hong Cao, Shao-Long Chen, Yandong Liu, Rui Zhang, and Ya~Zhang.
\newblock Limiting top quark-{Higgs} boson interaction and {Higgs}-boson width from multitop productions.
\newblock {\em Phys. Rev. D}, 99:113003, 2019.

\bibitem{CMS:TOP-22-009}
A.~Hayrapetyan et~al.
\newblock {Inclusive and differential cross section measurements of $\mathrm{t\bar{t}b\bar{b}}$ production in the lepton+jets channel at $\sqrt{s}$ = 13 TeV}.
\newblock Submitted to \textit{JHEP}, 2023.

\bibitem{Bols:2020bkb}
E.~Bols, J.~Kieseler, M.~Verzetti, M.~Stoye, and A.~Stakia.
\newblock Jet flavour classification using \textsc{DeepJet}.
\newblock {\em JINST}, 15:P12012, 2020.

\bibitem{LeCun:1989nips}
Yann LeCun, Bernhard~E. Boser, John~S. Denker, Donnie Henderson, Richard~E. Howard, Wayne~E. Hubbard, and Lawrence~D. Jackel.
\newblock Handwritten digit recognition with a back-propagation network.
\newblock In {\em {Proc. 2nd Int. Conf. on Advances in Neural Information Processing Systems (NIPS'89): Denver CO, USA, November 27--30, 1989}}, page 396, 1989.

\bibitem{Hochreiter:1997yld}
Sepp Hochreiter and J{\"u}rgen Schmidhuber.
\newblock Long short-term memory.
\newblock {\em Neural Comput.}, 9:1735, 1997.

\end{thebibliography}
\bibliographystyle{unsrt}
\end{multicols}
\end{document}